\documentclass[10pt,twocolumn,letterpaper]{article}

\usepackage{iccv}
\usepackage{times}
\usepackage{epsfig}
\usepackage{graphicx}
\usepackage{amsmath}
\usepackage{amssymb}

\usepackage{algorithm}
\usepackage[noend]{algpseudocode}

\usepackage{xcolor}
\usepackage{listings}

\definecolor{mGreen}{rgb}{0,0.6,0}
\definecolor{mGray}{rgb}{0.5,0.5,0.5}
\definecolor{mPurple}{rgb}{0.58,0,0.82}
\definecolor{backgroundColour}{rgb}{0.95,0.95,0.92}

\lstdefinestyle{CStyle}{
    backgroundcolor=\color{backgroundColour},   
    commentstyle=\color{mGreen},
    keywordstyle=\color{magenta},
    numberstyle=\tiny\color{mGray},
    stringstyle=\color{mPurple},
    basicstyle=\footnotesize,
    breakatwhitespace=false,         
    breaklines=true,                 
    captionpos=b,                    
    keepspaces=true,                 
    numbers=left,                    
    numbersep=5pt,                  
    showspaces=false,                
    showstringspaces=false,
    showtabs=false,                  
    tabsize=2,
    language=C
}

\usepackage{mdframed}
\usepackage{lipsum}
\newmdtheoremenv{theo}{Theorem}

\usepackage[breaklinks=true,bookmarks=false]{hyperref}

\iccvfinalcopy 

\def\iccvPaperID{****} 

\ificcvfinal\fi

\begin{document}

\title{Structure-based Optical Logics Without Using Transistors}

\author{Jonghyeon Lee\\
Sejong Academy of Science and Arts\\
Sejong, Korea\\
{\tt\small manatee02@sasa.hs.kr}
\and
Taewon Kang*\\
Department of Computer Science and Engineering\\
Korea University\\
Seoul, Korea\\
{\tt\small itschool@itsc.kr} \\
{* indicates corresponding author}
}


\maketitle
\ificcvfinal\thispagestyle{empty}\fi

\begin{abstract}
   The commercialization of transistors capable of both switching and amplification in 1960 resulted in the development of second-generation computers, which resulted in the miniaturization and lightening while accelerating the reduction and development of production costs. However, the self-resistance and the resistance used in conjunction with semiconductors, which are the basic principles of computers, generate a lot of heat, which results in semiconductor obsolescence, and limits the computation speed (clock rate). In implementing logic operation, this paper proposes the concept of Structure-based Computer which can implement NOT gate made of semiconductor transistor only by Structure-based twist of cable without resistance. In Structure-based computer, the theory of 'inverse signal pair' of digital signals was introduced so that it could operate in a different way than semiconductor-based transistors. In this paper, we propose a new hardware called Structure-based computer that can solve various problems in semiconductor computers only with the wiring structure of the conductor itself, not with the silicon-based semiconductor. Furthermore, we propose a deep-priority exploration-based simulation method that can easily implement and test complex Structure-based computer circuits. Furthermore, this paper suggests a mechanism to implement optical computers currently under development and research based on structures rather than devices.

\end{abstract}

\section{Introduction}
The structure-based computer mentioned in this paper are based on Boolean Algebra, a system commonly applied to digital computers. Boolean algebra is a concept created by George Boole (1815-1854) of the United Kingdom that expresses the True and False of logic 1 and 0, and mathematically describes digital electrical signals. The concept of logical aggregates defined in Boolean algebra has become the basis for hardware devices such as ALU, CLU, RAM, and so on. Structure-based computer in this paper was also designed to perform logical operations using digital signals of 1 and 0. Logic circuits are the units in which logical operations are performed, and there are AND, OR, and NOT gates. Of these, the NOT gate in the computer we use today is based on transistors. The advantage of transistors is that they can differentiate between signal and power and perform switching and amplification at the same time. On the other hand, more heat is generated compared to passing through a conductor of the same length, which causes semiconductors to age and limits the number of clocks. To solve the various problems of the semiconductor mentioned above, this paper shows the concept of "Reverse-Logic pair of digital signals" and "double-pair(4-pin)-based logic operation" techniques on which Structure-based computer hardware is. This paper shows the concept of Reverse-Logic pair\cite{phoneworks} of digital signals, which is a method for solving the problem of heating, aging, and computation speed of NOT operations. Expressing 1 as an inverted signal pair, it appears as an ordered pair of two auxiliary signals, each with a signal of one or zero, as shown in (1,0). Similarly, zeros are expressed in sequence pairs (0,1). \\

In other words, for any digital signal A, 
\begin{equation}
    A=(\alpha,\beta)\rightarrow\ |A|=\alpha,\ \ \ \ \beta= \sim \alpha\ \nonumber
\end{equation}

it can be expressed as (an expression) and $\alpha$ is defined as a true signal for logical A, $\beta$ as an inverted signal. Using this, the logical operation of two signals A and B is expressed as follows.

\begin{equation}
    for\ \forall A,B\in \{ \left(1,0\right),\left(0,1\right) \} \nonumber 
\end{equation}
\begin{align}
    NOT\ A &= \sim \left(\alpha_A,\beta_A\right) \nonumber \\
    &= \left(\sim \alpha_A, \sim \left(\sim \alpha_A\right)\right) \nonumber \\
    &= \left(\beta_A,\alpha_A\right) \nonumber \\
    A\ AND\ B &= \left(\alpha_A\land\alpha_B,\sim\left(\alpha_A\land\alpha_B\right)\right) \nonumber \\
    &=\left(\alpha_A\land\alpha_B,\sim\alpha_A\vee\sim\alpha_B\right) \nonumber \\
    &=(\alpha_A\land\alpha_B,\beta_A\vee\beta_B) \nonumber \\
    A\ OR\ B &= \left(\alpha_A\vee\alpha_B,\sim\left(\alpha_A\vee\alpha_B\right)\right) \nonumber \\
    &=\left(\alpha_A\vee\alpha_B,\sim\alpha_A\land\sim\alpha_B\right) \nonumber \\
    &=(\alpha_A\vee\alpha_B,\beta_A\land\beta_B) \nonumber
\end{align}

As shown in the above method, logical aggregates can be constructed with structural wiring if digital signals are computed in pairs of inverted signals. Especially for the NOT gate, you can twist the $\alpha$ line and the $\beta$ line once, making it much simpler to operate than a semiconductor-based transistor that uses a traditional semiconductor element and a pore. In addition, cables were measured in pairs rather than in pairs to enable serial connection when AND operations were performed. The $\alpha$ signal and $\beta$ signal have values of 0 and 1, depending on the connection state of the wire.
\begin{figure}[h]
\begin{center}
\includegraphics[width=4.6cm]{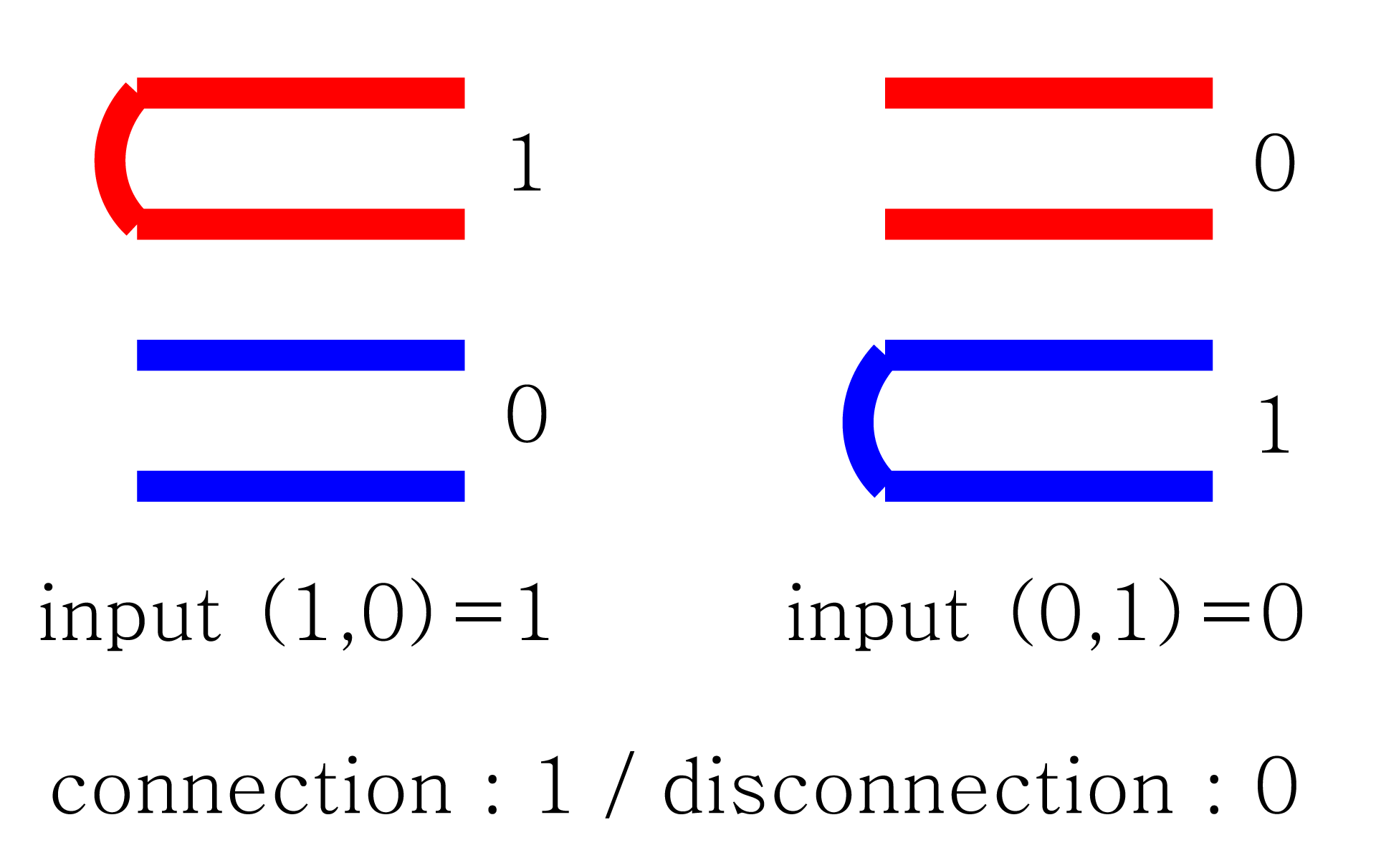}
\end{center}
\caption{\bf Double Pair-based Logical operation input method } \label{Fig01}
\end{figure}

If a pair of lines of the same color is connected, 1, if broken, the sequence pair of states of the red line ($\alpha$) and blue line ($\beta$) determines the transmitted digital signal. Thus, signal cables require one transistor for switching action at the end. When introducing the concept of an inverted signal pair of digital signals into a structural computer, the signals are paired, so a total of four wires are required to process the two auxiliary signals. This is defined as a double pair-based logical operation and is as follows in \textbf{Fig~\ref{Fig01}}.
\section{Methods}

\subsection{Structure-based Logic: 3-pin based logic}
The structural computer used an inverted signal pair to implement the reversal of a signal (NOT operation) as a structural transformation, i.e. a twist, and four pins were used for AND and OR operations as a series and parallel connection were required. However, one can think about whether the four pin designs are the minimum number of pins required by structural computers. In other words, operating a structural computer with a minimal lead is also a task to be addressed by this study because one of the most important factors in computer hardware design is aggregation. Let's look at the role of the four pins that transmit signals in a 4 pin based signal system. Four pins are paired into two pairs, each representing/delivering true and inverted values as a connection state. When checking the output, place a voltage on one of the two wires in a pair and ground the other. In this case, the study inferred that of the four wires, two wires acting as ground can be replaced by one wire, and based on this reasoning, the method in which the 4 pin signal system can be described as \textbf{3-pin based logic} as the same 3 pin signal system. As mentioned above, a 3-pin based logic consists of a ground cable in the center and two signal lines representing true and inverted values above and below, and is capable of operating NOT, AND and OR operations through the structural transformations shown below.

\begin{figure}[h]
\begin{center}
\includegraphics[scale=0.05,bb=0 0 600 328]{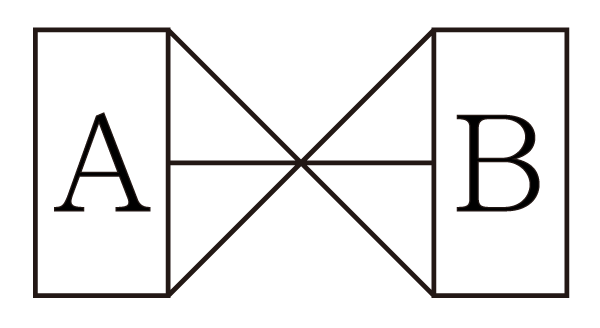}
\end{center}
\caption{\bf 3-pin based NOT gate}\label{Fig06}
\end{figure}

The NOT gate can be operated in a logic-negative operation through one ‘twisting’ as in a 4-pin. To be exact, the position of the middle ground pin is fixed and is a structural transformation that changes the position of the remaining two true and false pins.

\begin{figure}[h]
\begin{center}
\includegraphics[scale=0.2]{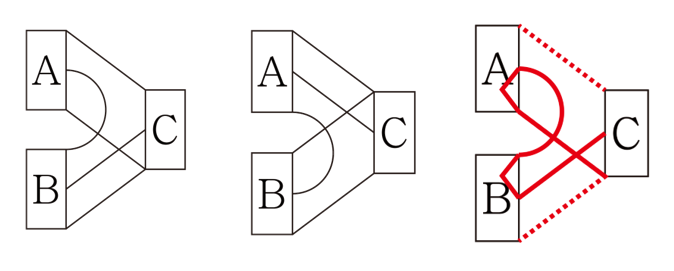}
\end{center}
\caption{\bf 3-pin based AND, OR, AND gate inputs A=0 and B=1}\label{Fig07}
\end{figure}



\textbf{Fig.~\ref{Fig07}} is AND and/or gate consisting of 3-pin based logics, \textbf{Fig.~\ref{Fig07}} also shows the connection status of the output pin when A=0, B=1 is entered in the AND gate. when A=0, B=1, or A is connected, and B is connected, output C is connected only to the following two pins, and this is the correct result for AND operation.

\section{Simulation with Depth-First Search}
To simulate the aforementioned structural computer theory, a device in the form of a USB connection. However, as the circuit grows in size, a number of USB-connected simulation devices are required, resulting in cost problems. We decided to verify that the structural computer theory presented so far is actually working without the cost of circuit building, to simulate the connection of complex Circuits rather than just Gate Circuit, and to set up Metric for experiments that can test structural computers for logical errors and error. 

\begin{figure}[t]
\begin{center}
\includegraphics[scale=0.1]{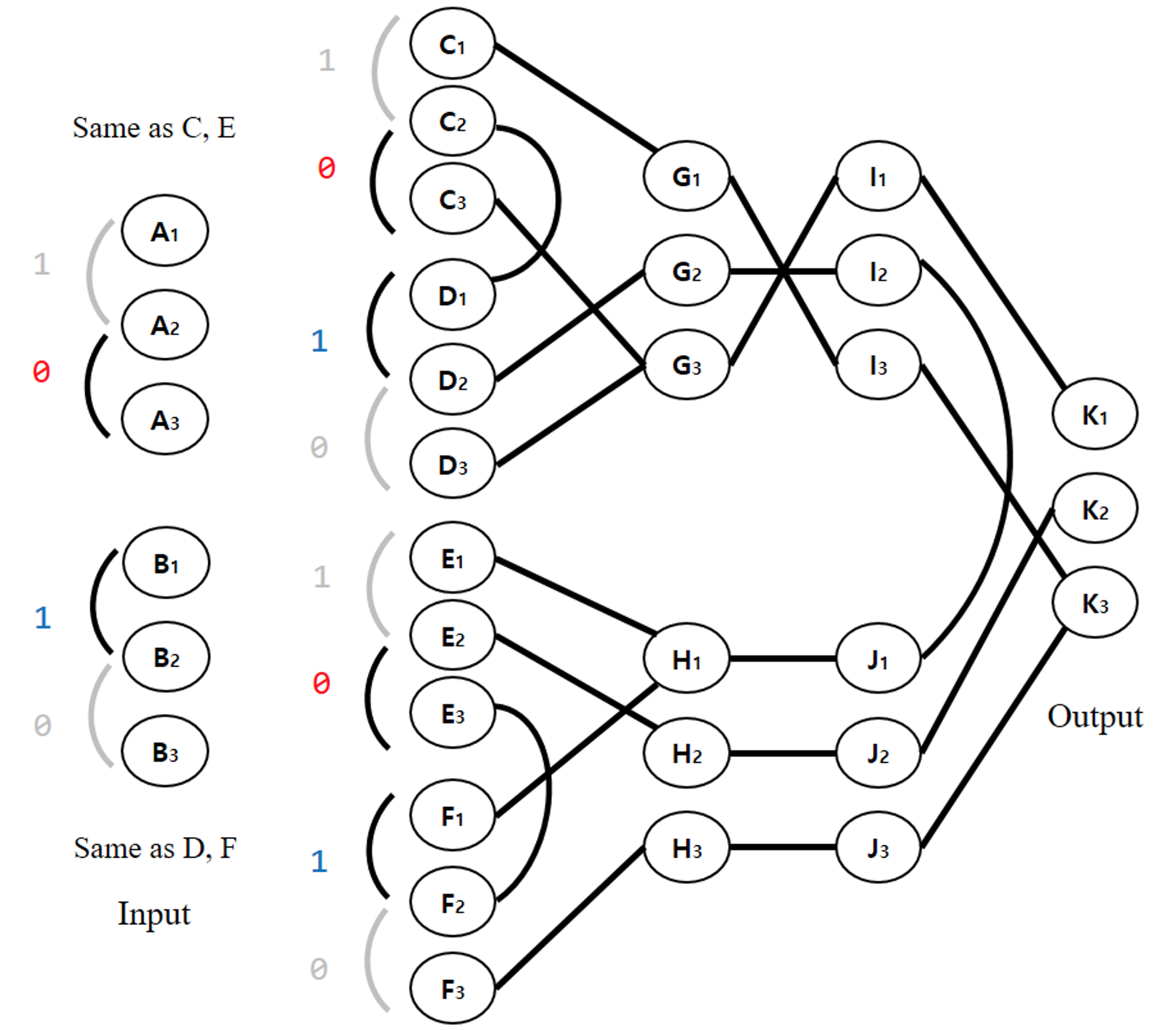}
\end{center}
\caption{\bf Example of XOR((A NAND B) AND (A OR B)) Gate into Vertex and Edge Graph}\label{dfs_nandor}
\end{figure}

\begin{table*}[h]
\centering
\begin{tabular}{|c|c|c|c|}
\hline
Start \& End & A & B & Result                                                \\ \hline
K2 $\rightarrow$ K1      & 0 & 0 & X                                                     \\ \hline
K2 $\rightarrow$ K1      & 1 & 0 & K2 J2 H2 E2 E1 H1 J1 I2 G2 D2 D3 G3 I1 K1                         \\ \hline
K2 $\rightarrow$ K1      & 0 & 1 & K2 J2 H2 E2 E3 F2 F1 H1 J1 I2 G2 D2 D1 C2 C3 G3 I1 K1 \\ \hline
K2 $\rightarrow$ K1      & 1 & 1 & X                         \\ \hline
K2 $\rightarrow$ K3      & 0 & 0 & K2 J2 H2 E2 E3 F2 F3 H3 J3 K3                         \\ \hline
K2 $\rightarrow$ K3      & 1 & 0 & X                                                     \\ \hline
K2 $\rightarrow$ K3      & 0 & 1 & X                         \\ \hline
K2 $\rightarrow$ K3      & 1 & 1 & K2 J2 H2 E2 E1 H1 J1 I2 G2 D2 D1 C2 C1 G1 I3 K3             \\ \hline
\end{tabular}
\caption{Result of Input and Output test case}
\label{dfs_inout_result}
\end{table*}

\subsection{DFS(Depth-First Search) Verification Simulation}
Graph described in \textbf{Fig. ~\ref{dfs_nandor}} is an implementation of an XOR gate combining NAND and OR, expressed in 33 vertices and 46 mains. Graphs are expressed in red and blue numbers in cases where there is no direction of the main line (the main line that can be passed in both directions) and the direction of the main line (the main line that can only be moved outward from the middle of the set of vertex). 


DFS (Depth First Search) verifies that the output is possible for the actual Pin connection state. As described above, the output is determined by the 3-pin input, so we will enter 1 with the A2 and A1 connections, the B2 and B1 connections (the reverse is treated as 0), and the corresponding output will be recognized through DFS navigation. In this course, we experiment with a total of eight test cases, including the number of input branches (four) of XOR and the direction of mobility of the output pin (K1 in K2 and K3 in K2).




\subsection{DFS (Depth First) Verification Simulation Results}

We will look at the inputs through 18 test cases to see if the circuit is acceptable. Next, it verifies with DFS that the output is possible for the actual pin connection state. As mentioned above, the search is carried out and the results are expressed by the unique number of each vertex. The result is as shown in \textbf{Table.~\ref{dfs_inout_result}}. The result of moving from the K2 peak to the K1 peak is the same as that of the XNOR, and the result of moving from the K2 peak to the K3 peak is the same as that of the XOR, it is possible to confirm that this study is feasible.

However, this circuit can confirm that circuit discovery errors occur in Y-shaped grinding (C3 to G3, D3 to G3 / E1 to H1 / I1 to K3, J1 to K3) because the electricity is unconditionally moving to low potential.

\section{Structure-based optical computer}
\subsection{Structure of Light-dependence and Geometric optics}
Exploration based on previous experiments and graph theory found errors in structural computers with electricity as a medium. The cause of these errors is the basic nature of electric charges: ‘flowing from high potential to low’. In short, the direction of current, which is the flow of electricity, is determined only by the height of the potential, not by the structure or shape of the circuit.
Unlike these biographies, however, light is so structural-dependent that there is geometrical optics, which is a study of the placement of mediums and their trajectory by their shape, which is straight forward by Fermat’s principle of minimum time. Thus, to address errors in electricity, structural computers will be used to implement optical computing.
Optical computing, which has been in the limelight for some time now due to its low heat output and fast computation, has been making progress. This is because it is difficult to physically implement optical devices to replace transistors, if they are built based on the theory of structural computers, they will be able to implement simple optical computing without the elements.

\subsection{Implementation of Structural Computer Using Mirrors and Translucent Mirrors}
Now, we will define ‘window operators’ to have the same connection as a 3-pin based structural computer using the reverse signal pair described earlier. ‘Window operator’ is a cube of 3x3, each containing elements of 0,i,1,-1,2, and 2. Each element (or cell) is inputted in the same way as three pin structural computing on the upper and lower surfaces. I will call it this because it is a basic unit that makes up an organization called a window operator. The expressions and functions of ) are as follows. \footnote{0, NULL: Transmits light that enters the upper and lower sides. i, Black body: absorbs the light from the top to the left. 1, Forward Mirror: Double-sided mirror with a 45 degree inclination -1, Reverse Mirror: Double-sided mirror perpendicular to 1. 2, forward half-mirror: A translucent object with a 45 degree gradient, some light is transmittable, some reflective. -2, Reverse Semi-mirror: A translucent object perpendicular to 2. } Based on the above functions, the window operator is designed to allow the same operation of the connected state as the three pin based AND gate shown earlier.

\begin{figure}[h]
\begin{center}
\includegraphics[scale=0.35]{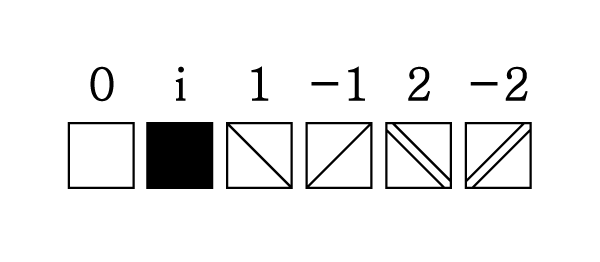}
\end{center}
\caption{\bf The cells that make up the window operator (organization).
All window operators shall base their left-to-right and top-down directions forward and vice versa.}\label{421}
\end{figure} 
 
\begin{figure}[h]
\begin{center}
\includegraphics[scale=0.1]{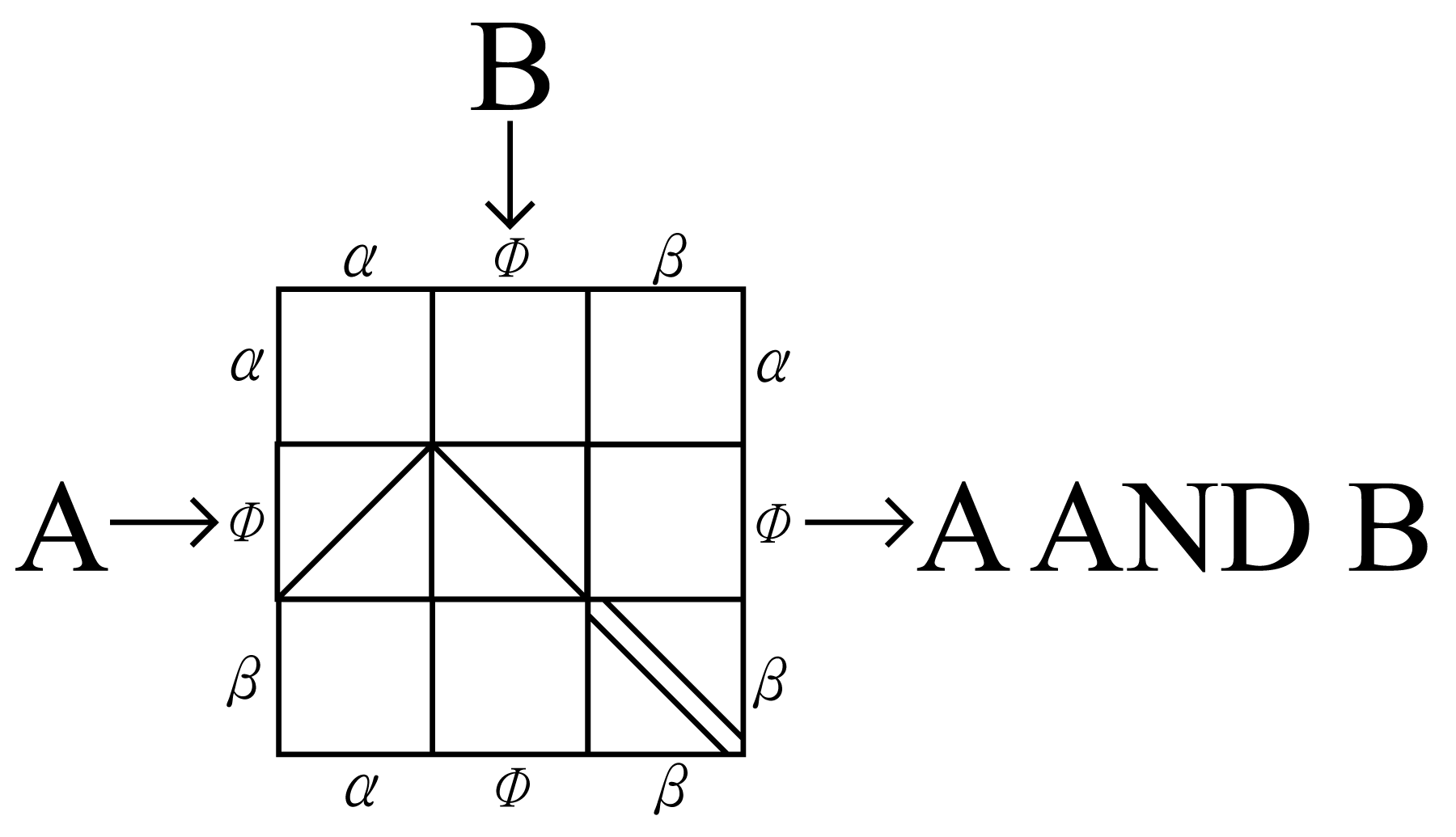}
\end{center}
\caption{\bf Window Operator Representation of AND Gate.}\label{422}
\end{figure} 

This window operator calculates the connection between the pie and alpha, or beta, at A and B and transfers it to the right side (A AND B). In case of output, it is possible to measure by firing a laser onto a pie pin on the resulting side and checking whether it returns to either alpha or beta. The picture shows the connection status determination of the results for each input.
  
\begin{figure}[h]
\begin{center}
\includegraphics[scale=0.1]{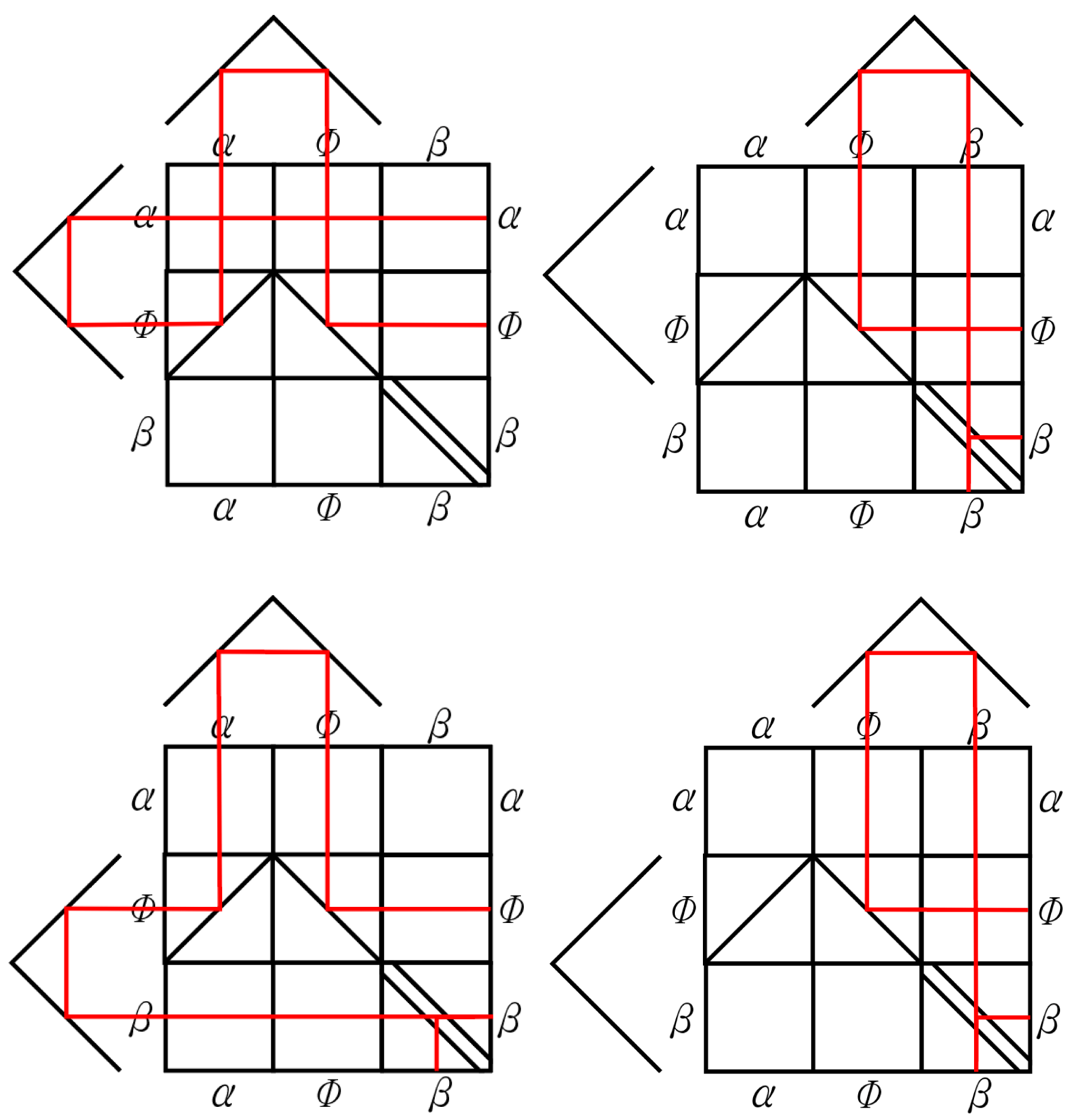}
\end{center}
\caption{\bf Output verification according to all inputs of the AND gate: 1 AND 1=1, 1 AND 0=0, 0 AND 1 AND 1=0, 0 AND 0=0, respectively, from the top left, present the correct output value from the AND gate and all cases.}\label{423}
\end{figure} 

\subsection{Multifunctional Expression and Function of Structural-Based Optical Non-Grid}
Optical logic aggregates can be designed in the same way as in \textbf{Implementation of Structural Computer Using Mirrors and Translucent Mirrors}, and for the convenience of expression and the exploration of mathematical properties (especially their association with matrices), the number shown in \textbf{Fig.~\ref{421}} can be applied to the label of the window operator to express the AND gate as shown below, which is referred to as the matrix representation of the optical logic. \textbf{Fig.~\ref{423}} shows, however, that some rays of light can be counted on the lower beta signal, which can interfere with the operation of other Thus, a black body gate was implemented using i cells to make input everywhere into NULL state. Including this, functions derived from the properties of light that are only available in structural-based optical computing can be modularized with window operators, which can be organized into the following seven categories. \footnote{AND- Logic in Boolean algebra, OR- Logic in Boolean algebra, CROS- Vertical Reflection/Crossing of Two Logics, CNOT- Vertical Reflection/Crossing of Two Logics, Only Intersects and Both Logics are NOT-operated. INVS- Transmittance of Two Logics, COPY- Cloning Logic, BLAK- Absorption of logic (to make it all NULL) }

\begin{figure}[h]
\begin{center}
\includegraphics[scale=0.1]{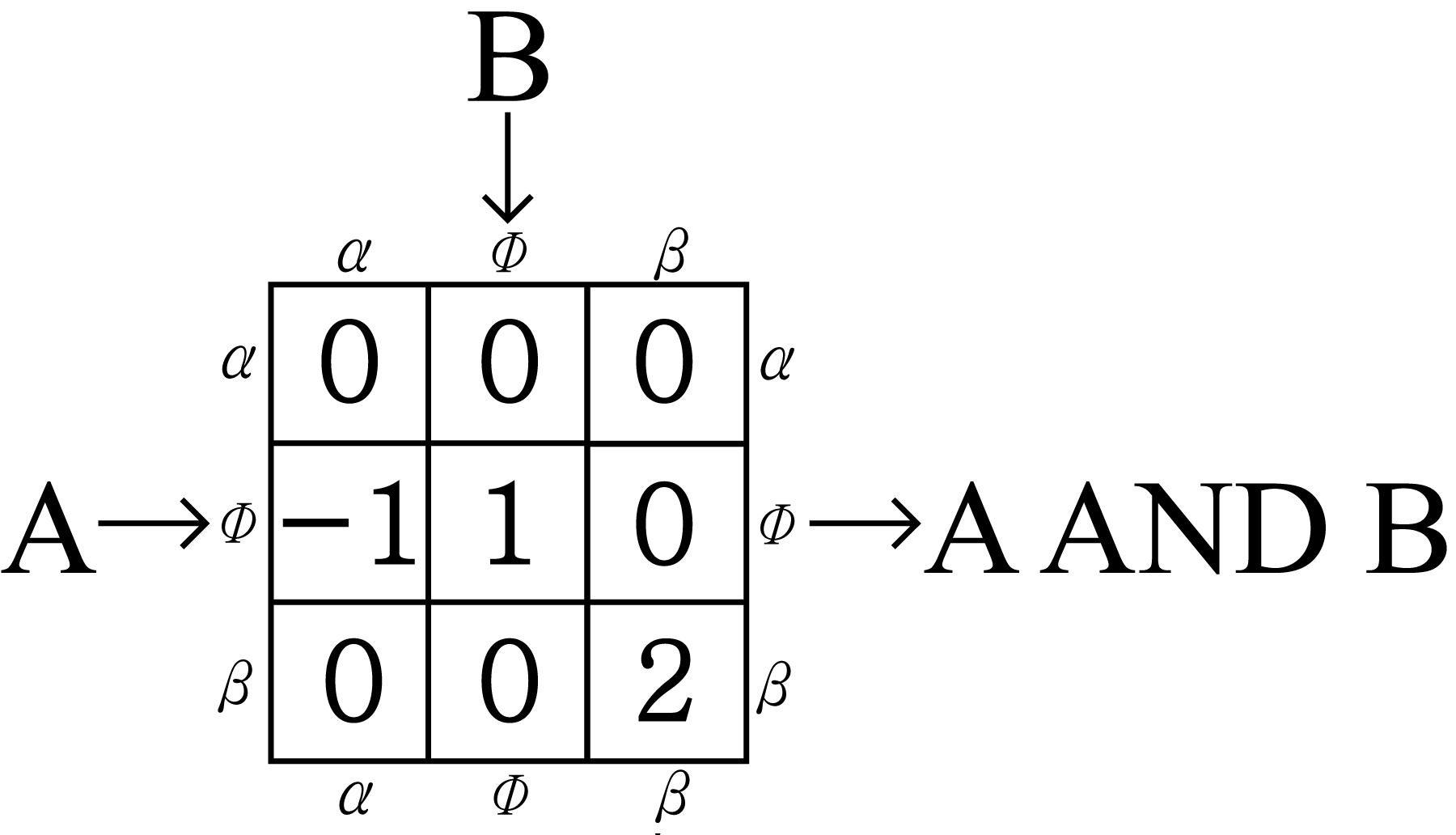}
\end{center}
\caption{\bf Marking of the AND gate matrix.}\label{431}
\end{figure} 

\begin{figure}[h]
\begin{center}
\includegraphics[scale=0.12]{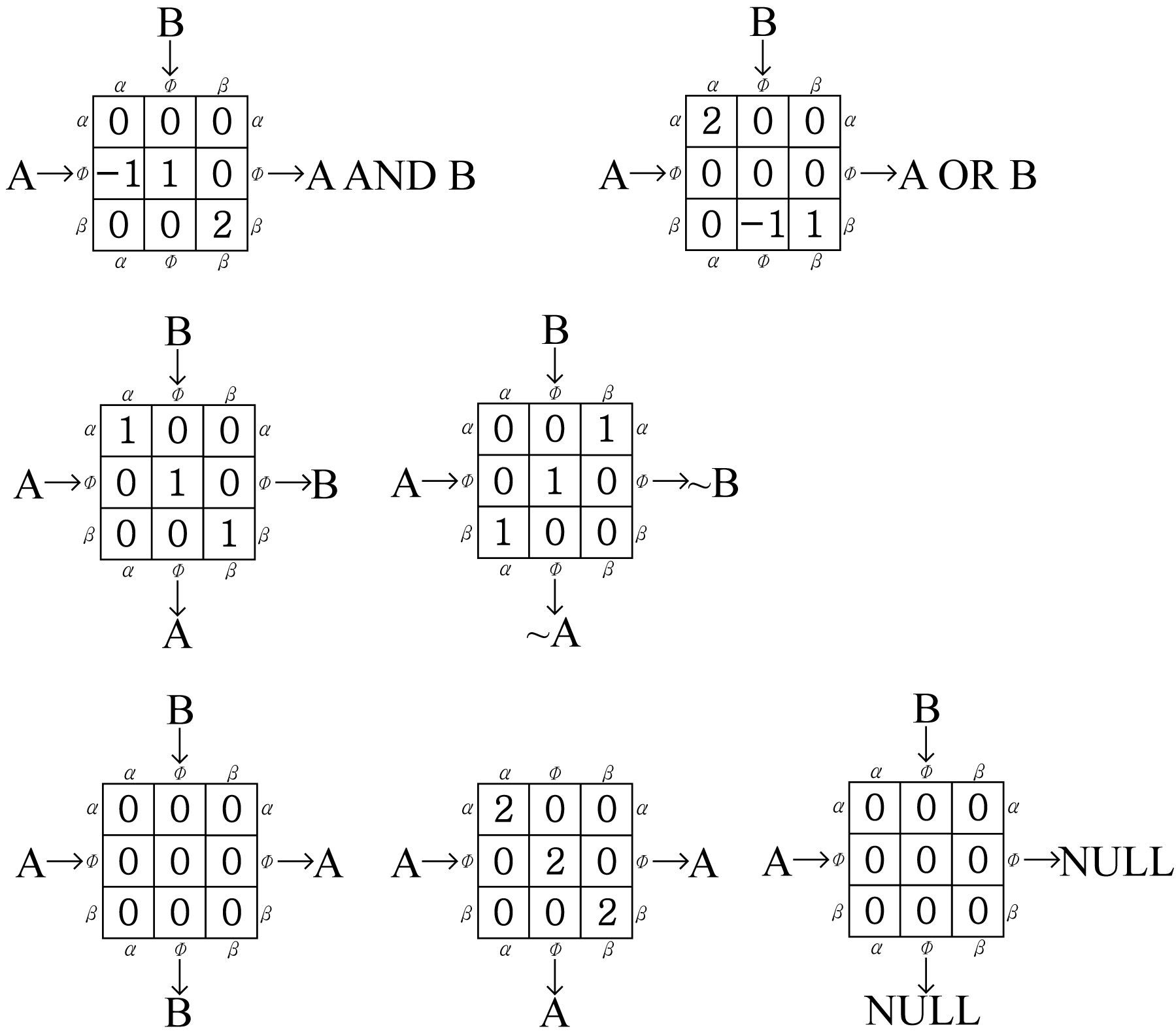}
\end{center}
\caption{\bf Matrix representation of basic optical logic aggregates (or window operators): AND, OR, CROS (cross), CNOT (cross-not), INVS (invisible), COPY (back-body) gates, and BLAK (back-body) gates, starting from the top left.}\label{432}
\end{figure}

\section{Conclusion}
This paper presents the NOT gate implementation of structural computers and the Reverse-Logic pair and double pair-based logic operation techniques of digital signals that can solve the problem of heating and aging of existing semiconductor computers. 



Furthermore, we propose Simulation Metric (DFS) based on deep-first search (DFS) that enables easy implementation and testing of complex structural computer Circuits. This confirmed the feasibility of this study in an experiment based on an XOR gate produced by combining NAND, AND and OR gates.


And it is expected that this research can be applied to the development of artificial intelligence technologies such as deep learning in the future. In other words, it is expected that the idea of structural computers will be applied to semiconductors that generate a lot of heat, such as Computer Vision task\cite{gpu}\cite{gpu1}\cite{gpu2}\cite{gpu3}\cite{gpu4}\cite{gpu5}\cite{gpu6}\cite{gpu7}\cite{gpu8} that require GPU processing of large amounts of data, to drastically reduce heat generation, reduce electricity use, and improve performance more than before.


\end{document}